\documentclass[preprint,review,12pt]{elsarticle}

\usepackage{amssymb}
\usepackage{moreverb}
\usepackage{epstopdf}
\usepackage{booktabs}
\usepackage{lscape}
\usepackage{bbm}
\usepackage{url}

\journal{arXiv}

\begin{document}

\def\spacingset#1{\renewcommand{\baselinestretch}%
{#1}\small\normalsize} \spacingset{1.45}

\begin{frontmatter}


\title{\Large {\bf Spatio-Temporal Forecasting by Coupled Stochastic Differential Equations: Applications to Solar Power}}

\author[address1]{ Emil B. Iversen}
\author[address1]{ Rune Juhl}
\author[address1]{ Jan K. M\o ller}
\author[address2]{ Jan Kleissl}
\author[address1]{ Henrik Madsen}
\author[address1]{ Juan M. Morales}

\address[address1]{Technical University of Denmark, Asmussens Alle, building 322, DK-2800 Lyngby, Denmark}
\address[address2]{University of California, San Diego, 9500 Gilman Drive MC 0411, La Jolla, CA 92093, USA}


\begin{abstract}

Spatio-temporal problems exist in many areas of knowledge and disciplines ranging from biology to engineering and physics. However, solution strategies based on classical statistical techniques often fall short due to the large number of parameters that are to be estimated and the huge amount of data that need to be handled. In this paper we apply known techniques in a novel way to provide a framework for spatio-temporal modeling which is both computationally efficient and has a low dimensional parameter space. We present a micro-to-macro approach whereby the local dynamics are first modeled and subsequently combined to capture the global system behavior. The proposed methodology relies on coupled stochastic differential equations and is applied to produce spatio-temporal forecasts for a solar power plant for very short horizons, which essentially implies tracking clouds moving across the field of solar power inverters. We outperform simple and complex benchmarks while providing forecasts for 70 spatial dimensions and 24 lead times (i.e., for a total number of random variables equal to 1680). The resulting model can provide all sorts of forecast products, ranging from point forecasts and co-variances to predictive densities, multi-horizon forecasts, and space-time trajectories.

\end{abstract}


\end{frontmatter}
\newpage
\spacingset{1.45}

\section{Introduction} \label{sec: Intro}

Recent years have seen a massive increase in the collection of data. The challenges related to the treatment of large datasets have, in turn, been the subject of intense research. Among these challenges are the development of predictive methods for spatio-temporal systems. In this article we are concerned with modeling the spatio-temporal behavior in order to better predict the aggregate system dynamics. The need for high dimensional predictions arises in many fields. In meteorology and hydrology such methods allow for precipitation nowcasting to predict flooding and to predict extreme wind speeds (\cite{versini2012use}, \cite{sigrist2014stochastic}, \cite{xu2005kernel}, \cite{moller2013multivariate}). Applications in the field of biology have ranged from the spread of diseases to models for genomics (\cite{van2003modelling}, \cite{robertson2010review}, \cite{waller1997hierarchical} \cite{richardson2010bayesian}, \cite{bowman2007spatiotemporal}). In social sciences spatio-temporal models can be used to predict the behavior and response of individuals and groups (\cite{lazer2009life}). An abundance of further examples and references for spatio-temporal analyses can be found in \cite{cressie2011statistics}.

In general, there are two main modeling approaches to spatio-temporal problems: those driven by the underlying physics of the system (deterministic) and those driven by data (stochastic). Examples of physically driven systems are models for producing numerical weather predictions (\cite{uppala2005era}), describing the dynamics of a boiler (\cite{aastrom2000drum}) and modeling mud flow down a slope (\cite{huang1998herschel}). These types of models have their roots in a set of physical laws or local behavior such as the conservation of momentum or mass and energy balances. Examples of data-driven models include modeling space-time scenarios of wind power generation (\cite{tastu2013space}), mapping of disease rates (\cite{waller1997hierarchical}) and modeling of sea surface temperature (\cite{lemos2009spatio}). These approaches share the emphasis on the aggregate system behavior as opposed to its local dynamics and tackle the modeling task by placing the focus on probability distributions, correlations and inter-dependences.

This article adds to the literature on spatio-temporal modeling by providing a data-driven approach to capture the local dynamics of a larger system. Specifically, we initially propose a model for the local dynamics using coupled stochastic differential equations. This model is, in turn, generalized to govern all local behavior, thus yielding a global model. Furthermore, this global model can be interpreted as a discretization of a more general model in space and time given by a system of partial stochastic differential equations.

\subsection{A Motivating Example} \label{sec: Intro Example}

The practical application that motivates our methodology is that of forecasting the power generated by a solar power plant. Our claim is that, by understanding and capturing the specific spatio-temporal dynamics of the solar field, we will be able to improve our power predictions. The power output of a solar plant exhibits dynamics on several different time scales: from cycles spanning a year governed by the sun height and climate, to dynamics in the range of weeks and days governed by weather fronts, to hourly dynamics governed by local weather phenomena and different sun height during the day, to the very short horizon of minutes and seconds as a result of the movement of individual clouds. In this example we are motivated by nowcasting (forecasting in the range from seconds to minutes) the power output of a solar plant by tracking the movement of individual clouds across the solar field.

\begin{figure}[!ht]
\begin{center}
  \includegraphics[width=0.90\textwidth]{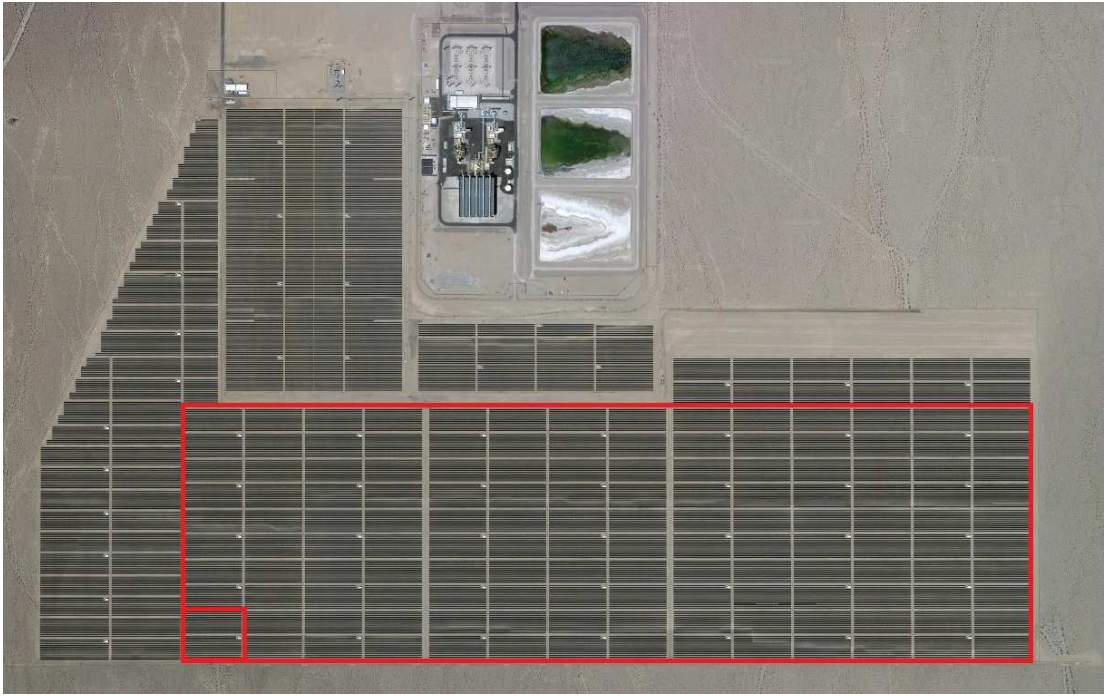}\\
    \vspace{-10pt}
  \caption{\emph{An aerial photo of the Copper Mountain Solar Facility. The small red rectangle indicates the size of a single inverter, which is 125m by 125m. The large red rectangle indicates the 5 by 14 inverters for which we provide forecasts of the power production. }}\label{fig:CopperMountain2}
\end{center}
\vspace{-10pt}
\end{figure}

Our data pertain to the Copper Mountain Solar 1 Facility with a rated capacity of 58 MW, and part of the Copper Mountain Solar Facility with at total rated capacity of 150 MW. Our measurements stem from 96 photovoltaic inverters, each with measurements taken every second. We limit the study to the rectangular grid of 5 by 14 inverters shown in red in Figure \ref{fig:CopperMountain2}. This specific cutout data is also used in the paper by \cite{lipperheide2015embedded}.

Good forecasts for power production are extremely important for secure, reliable, and efficient operation of the electrical grid \citep{pinson2013wind, morales2014decision}. Forecasting the power output dynamics on very short horizons has proven essential for an efficient integration of solar power. On partly cloudy days the power output of a solar plant can drop from nominal capacity to between 20-25\% of nominal capacity within just a minute as a cloud passes overhead. With a nominal capacity of 150 MW, a drop of 75\% results in a power output decrease of 112.5 MW, which may severely challenge grid stability. Very short-term forecasts may help mitigate the detrimental effects of this power drop by either installing storage devices or by providing an early warning for grid operators. Spatio-temporal forecasts of solar irradiance have been an output of physical models such as numerical weather prediction models for decades (\cite{uppala2005era}). However, besides being deterministic in nature, numerical weather prediction models do not provide enough resolution in space or time to track clouds. Spatio-temporal models for solar power production have just recently received attention: \cite{yang2014multitime} formulate an auto-regressive time-series model for capturing the correlation in the solar power output across a small area. In \cite{yang2014solar} space-time kriging and a vector auto-regressive model are employed to describe the spatio-temporal solar power production. \cite{lipperheide2015embedded} outperform a persistence model for short horizons by means of a cloud speed persistence model that propagates the solar power production across the spatial grid in the direction of the cloud speed.

Using stochastic differential equations to model physical systems with a large random component is not new. However, the coupling and modeling methodology presented here is a novel combination and it constitutes a step forward for spatio-temporal modeling in terms of computational efficiency, a low-dimension parameter space, predictions for multiple horizons and characterization of the spatio-temporal interdependence. Furthermore, for the specific case of solar power forecasting, the proposed modeling approach allows for a particularly elegant interpretation of the global system dynamics. The rest of the paper is structured as follows: Section \ref{sec: SDEs} introduces the proposed stochastic differential equation framework. Section \ref{sec: Spatio-Temporal Models} describes the approach for using this framework to develop spatio-temporal models in general and for the specific application to solar power forecasting. In Section \ref{sec: SPDE interpretation } we obtain a generalized interpretation of the proposed model. Section \ref{sec: CTSM-R} details the parameter estimation procedure. We then evaluate the performance of the obtained spatio-temporal model on a real-world example in Section \ref{sec: Nowcasting at Copper Mountain}. Lastly, Section \ref{sec: Conclusion} concludes the paper.

\section{Stochastic Differential Equations} \label{sec: SDEs}

Stochastic differential equations (SDEs) are an extension of ordinary differential equations (ODEs) obtained by including one or more stochastic terms. The solution to a SDE is a stochastic process describing the evolution of a random variable over time. SDEs have been used to describe a variety of phenomena governed by a large random component and are especially prominent in finance (\cite{bjork2009arbitrage}, \cite{mikosch1998elementary}) and physics (\cite{van1992stochastic}, \cite{adomian1988nonlinear}). We give here a very short introduction to SDEs and refer the interested reader to \cite{oksendal2010stochastic} for a complete and thorough treatment of the subject.

Suppose that we have a continuous time process $U_t \in \mathcal{U} \subset \mathbb{R}^n$. From ordinary differential equations the evolution in time of the state variable, $U_t$, is defined by the deterministic system equation:
\begin{eqnarray}
\frac{dU_t}{dt} = f(U_t,t),
\end{eqnarray}
where $t \in \mathbb{R}$ and $f(\cdot) \in \mathbb{R}^n$. For complex systems the dynamics may be too intricate to be captured fully by $f(\cdot)$ or there may be random perturbations of inputs that are not specified by the model. This suggests the introduction of a random component in the state evolution to capture such perturbations or model deficiencies. By introducing a random component in the dynamics of the state process, as carried out in \cite{oksendal2010stochastic}, we obtain the following state process:
\begin{eqnarray}
\frac{dU_t}{dt} = f(U_t,t) + g(U_t,t) W_t, \label{eq: SDE formulation 1}
\end{eqnarray}
where $W_t \in \mathbb{R}^m$ is an $m$-dimensional standard Wiener process and $g(\cdot) \in \mathbb{R}^{n \times m}$ is a matrix function. Multiply by $dt$ on both sides of (\ref{eq: SDE formulation 1}) to obtain the standard SDE formulation:
\begin{eqnarray}
dU_t = f(U_t,t) dt + g(U_t,t) dW_t. \label{eq: SDE formulation 2}
\end{eqnarray}
This standard formulation for SDEs is not well defined, as the derivative of $W_t$, $\frac{dW_t}{dt}$, does not exist. Instead, equation (\ref{eq: SDE formulation 2}) should be interpreted as an informal way of writing the integral equation:
\begin{eqnarray}
U_t = U_0 + \int_0^t f(U_t,t) dt + \int_0^t g(U_t,t) dW_s. \label{eq: SDE formulation 3}
\end{eqnarray}
In equation (\ref{eq: SDE formulation 3}) the behavior of the stochastic process $U_t$ is expressed as the sum of an initial stochastic variable, a Lebesgue integral and an It\={o} integral, respectively.

In general, it is only feasible to observe a continuous time process in discrete time. To this end, we observe the process $U_t$ at discrete times through an observation equation. Let $Y_k \in \mathcal{Y} \subset \mathbb{R}^l$ denote the observation at the discrete time $t_k$. We define the observation equation as:
\begin{eqnarray}
Y_k = h \left( U_{t_k}, t_k, e_k \right),
\end{eqnarray}
where the introduction of $t_k$ allows for some external input, $e_k \in \mathbb{R}^l$ is the observation error and the function $h(\cdot) \in \mathbb{R}^l$ links the process state to the observation.

 The solution to a deterministic ordinary differential equation is a point for each future time $t$. In the SDE setting the solution is a stochastic process with a probability density for any state and for any future time $t$. For an It\={o} process defined as in (\ref{eq: SDE formulation 3}) with drift $f(U_t,t)$ and $g(U_t,t) = \sqrt{2 D(U_t,t)}$, the probability density function $p(u,t)$ in state $u$ at time $t$ of the random variable $U_t$ is given as the solution to the partial differential equation known as the Fokker-Planck equation (\cite{bjork2009arbitrage}):
\begin{eqnarray}
\frac{\partial}{\partial t} p(u,t) = - \frac{\partial}{\partial u}\left[ f(u,t) p(u,t) \right] + \frac{\partial^2}{\partial u^2}\left[ D(u,t) p(u,t) \right].
\end{eqnarray}
Thus, given a specific SDE formulation, the predictive density for any future time can be obtained by solving a partial differential equation. While analytic solutions only exist for particularly simple SDE formulations, a host of numerical solutions are available (\cite{johnson2012numerical}, \cite{smith1965numerical}). While not solving the Fokker-Planck equation directly, another technique, the Monte Carlo approach, solves this problem implicitly by approximating the predictive density through simulation (\cite{robert2004monte}).

\section{A Spatio-Temporal Model by Coupled SDEs} \label{sec: Spatio-Temporal Models}

Consider a stochastic process in space $x$ and time $t$, and denote this process by $U(x,t)$. Suppose that there are $I \times J$ locations at $x_{i,j}$ where we want to model the process. First, let $U(x,t)$ at location $x_{ i,j }$ be denoted by $U(x_{i,j},t) = U_{i,j,t}$. Now suppose that we want to use a stochastic differential equation to represent the dynamics of each $U_{i,j,t}$. This gives us a model of the following form:
\begin{eqnarray}
dU_{i,j,t} &=& f \left( \mathbf{U}_t, t \right)dt + g( \mathbf{U}_t, t ) dW_{i,j,t} \label{eq: general model eq1} \\
Y_{l,k} &=& h( \mathbf{U}_{t_k}, t_k ) +  \epsilon_{l,k}, \label{eq: general model eq2}
\end{eqnarray}
where we let $\mathbf{U}_t$ be the vector containing all $U_{i,j,t}$ for a specific $t$.

Next, we enforce that two locational processes have a direct interaction only if they are adjacent to each other. This stems from a physical interpretation of the system, whereby we allow only locations that are in direct contact to interact with each other. To this end, define $\mathbf{U}_{i,j,t}$ as the set of $U_{i,j,t}$'s that are in the nearest neighborhood of $U_{i,j,t}$, thus $\mathbf{U}_{i,j,t} = \left\{ U_{i,j,t}, U_{i-1,j,t}, U_{i+1,j,t}, U_{i,j-1,t}, U_{i,j+1,t} \right\}$. Furthermore, assume that for each $U_{i,j,t}$ we allow $g(\cdot)$ to depend only on $U_{i,j,t}$ and $t$. This also follows from a physical interpretation of the system dynamics, whereby random perturbations at location $x_{ i,j }$ can only affect adjacent locations by first affecting $U_{i,j,t}$. This leads to the following model formulation:
\begin{eqnarray}
dU_{i,j,t} &=& f\left(\mathbf{U}_{i,j,t},t \right)dt + g(U_{i,j,t}, t) dW_{i,j,t} \label{eq: semi-general model eq1} \\
Y_{l,k} &=& h( \mathbf{U}_{t_k}, t_k ) +  \epsilon_{l,k}. \label{eq: semi-general model eq2}
\end{eqnarray}

\begin{figure}[!ht]
\begin{center}
    \vspace{-10pt}
  \includegraphics[width=0.6\textwidth,natwidth=200,natheight=200]{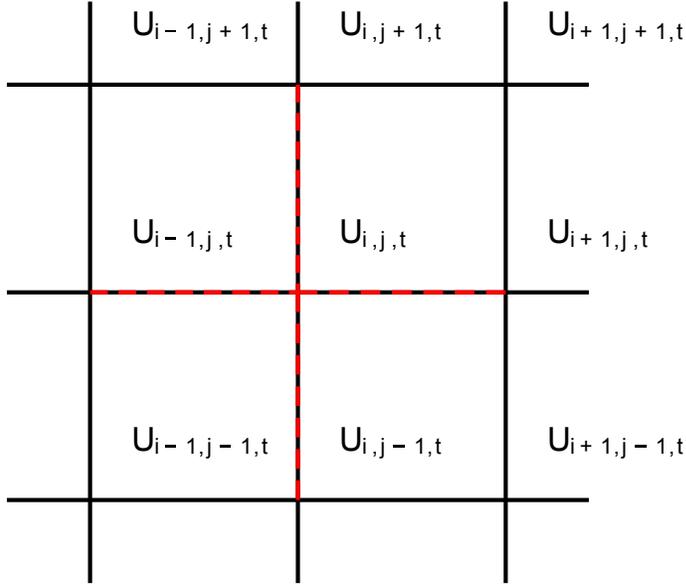}\\
    \vspace{-10pt}
  \caption{\emph{A stencil showing the relationship between $U(x,t)$ at different locations, with black lines indicating a direct relationship. The red dashed lines show the interactions for $U_{i,j,t}$}}\label{fig:DrawStencilSolar}
\end{center}
\vspace{-20pt}
\end{figure}

Since the model is formulated in continuous time, by appealing to the physical nature of the system, one should expect there to be no interaction between locations that are not adjacent to each other. The model formulation is illustrated in Figure \ref{fig:DrawStencilSolar}. Here we have depicted the different locations by dots and interactions with lines connecting two dots. Specifically we have highlighted in red all interactions that concern location $U_{i,j,t}$. This specific formulation is defined for 2-dimensions but can be easily generalized to higher dimensions.

\subsection{Application to Solar Power Forecasting} \label{sec: Solar Forecasting Model}

Based on the model formulation outlined in this section, we propose a model for predicting the power output of a photovoltaic solar power facility. The model exploits the power output of the up-wind solar inverters to predict the future power output of the down-wind solar inverters.

The data available pertains to the Sempra US Gas \& Power Copper Mountain Solar Facility outside Boulder City, Nevada, USA. The power output of the solar power inverters is normalized both with respect to the sun height and with respect to the solar panel tilt and we henceforth refer to the normalized power as power. We let the \emph{change} in power output of inverter $\left[ i,j \right]$ at location $x_{i,j}$ at time $t$ be modeled by the stochastic variable $U_{i,j,t}$. We order the inverters such that inverter $U_{i+1,j,t}$ is the one directly to the east of inverter $U_{i,j,t}$. Also, we name the inverters such that $U_{i,j+1,t}$ is the inverter directly north of $U_{i,j,t}$ (see Figure \ref{fig:DrawStencilSolar}).

 We have cloud speed measurements at our disposal, provided by the approach given in \cite{bosch2013cloud}. The cloud speed is denoted by $v_t$. For modeling purposes, the cloud speed is decomposed into its four directional components, namely, North, East, South, and West, denoted by $\textrm{n}_t$, $\textrm{e}_t$, $\textrm{s}_t$, and $\textrm{w}_t$, respectively. We employ four directional components instead of two, as we impose the condition that the directional component must be non-negative. Thus, a wind from South-East has positive South and East components, but zero North and West components. This way we end up with the model:
\begin{eqnarray}
dU_{i,j,t} &=&\theta | v_t | \Bigg( \textrm{n}_t(U_{i,j+1,t} - U_{i,j,t}) \mathbbm{1}_{\left\{j+1 \leq J \right\} } + \textrm{e}_t(U_{i+1,j,t} - U_{i,j,t})\mathbbm{1}_{ \left\{i+1 \leq I \right\} }  \nonumber \\
 && \left. + \textrm{s}_t(U_{i,j-1,t} - U_{i,j,t})\mathbbm{1}_{ \left\{ j-1 \geq 1 \right\} } + \textrm{w}_t(U_{i-1,j,t} - U_{i,j,t})\mathbbm{1}_{ \left\{ i-1 \geq 1 \right\} } \right. \\ \label{eq: SPDE model eq1}
&& - \mu U_{i,j,t} \left( \textrm{n}_t \mathbbm{1}_{ \left\{j = J \right\} } + \textrm{s}_t \mathbbm{1}_{ \left\{j = 1 \right\} } + \textrm{e}_t \mathbbm{1}_{ \left\{i = I \right\} } + \textrm{w}_t \mathbbm{1}_{ \left\{i = 1 \right\} } \right)  \Bigg) dt \nonumber \\ 
&& + \sigma dW_{i,j,t}  \nonumber \\ 
dQ_{i,j,t} &=& U_{i,j,t}dt \label{eq: SPDE model eq2} \\
Y_{i,j,k} &=& Q_{i,j,t_k} +  \epsilon_{i,j,k}. \label{eq: SPDE model eq3}
\end{eqnarray}
Here $Y_{i,j,k}$ is the observed power produced from location $x_{ i,j }$ at time $t_k$. $Q_{i,j,t}$ can thus be interpreted as the actual produced power at this location. This implies that $U_{i,j,t}$ is the change in power production at location $x_{ i,j }$ at time $t$. Notice that the spatial dynamics are modeled by equation (\ref{eq: SPDE model eq1}) and that equations (\ref{eq: SPDE model eq2})--(\ref{eq: SPDE model eq3}) correspond to the observation equation (\ref{eq: semi-general model eq2}). Thus, to express the SDE model~(\ref{eq: SPDE model eq1})--(\ref{eq: SPDE model eq3}) in the form of (\ref{eq: general model eq1})--(\ref{eq: general model eq2}), it suffices to write the $h(\cdot)$ function as  $h_{i,j}(t_k) = \int_{t_{k-1}}^{t_{k}} U_{i,j,s} ds + \epsilon_{i,j,k}$. Further we have that $\epsilon_{i,j,k} \sim \mathcal{N} \left( 0,\sigma^2_{\epsilon} \right)$. The parameters in the model are thus $\theta, \mu, \sigma$ and $\sigma_{\epsilon}$, where $\theta | v_t |$ governs the speed at which the value in adjacent cells tend towards each other. Parameter $\mu$ governs how rapidly the change in power output of inverter $\left\{ i,j \right\}$, $U_{i,j,t}$, tends to zero, if $U_{i,j,t}$ is an upwind cell. $\sigma$ is the system noise and $\sigma_{\epsilon}$ characterizes the observation noise. Symbol $\mathbbm{1}_{\left\{ \cdot \right\}}$ represents an indicator or heavyside function, that is equal to 1 if the stated condition is met, and 0 otherwise. The indicator functions are used to handle the boundaries of the solar field, such that the model only relates locations that are actually present in the model. This also applies to the dampening term, where we dampen cells on the leading edge towards the wind. Further, note that, in this particular case, $Y_{l,k} = Y_{i,j,k}$, where we let $l$ go through all the feasible combinations of $\left\{ i,j \right\}$ to conform with the notation in equation (\ref{eq: general model eq2}) (that is, model~(\ref{eq: SPDE model eq1})--(\ref{eq: SPDE model eq3}) assumes that we have power measurements for all locations or inverters).



\section{Continuous Space Interpretation} \label{sec: SPDE interpretation }

Given the model formulation (\ref{eq: SPDE model eq1})--(\ref{eq: SPDE model eq3}) one might ask what would happen if the grid size approached zero. Notice that, since we have a fixed distance $\Delta x$ between all adjacent grid points, we can normalize the model parameters by doing $ \theta =  \tilde{\theta} / \Delta x $, $ \sigma = \tilde{\sigma} \Delta x $ and $\mu = \tilde{\mu} \Delta x$, with $\Delta x = c$, where $c$ is some constant. Consequently, the SDE model (\ref{eq: SPDE model eq1})--(\ref{eq: SPDE model eq3}) can be recast as:
\begin{eqnarray}
dU_{i,j,t} &=& \tilde{\theta}  |v_t|  \left( \textrm{n}_t \left( \frac{U_{i,j+1,t} - U_{i,j,t}}{\Delta x} \right) \mathbbm{1}_{\left\{j+1 \leq J \right\} } + \textrm{e}_t \left( \frac{U_{i+1,j,t} - U_{i,j,t}}{\Delta x} \right) \mathbbm{1}_{ \left\{i+1 \leq I \right\} }  \right. \nonumber \\
 && \left. + \textrm{s}_t \left( \frac{U_{i,j-1,t} - U_{i,j,t}}{\Delta x} \right) \mathbbm{1}_{ \left\{ j-1 \geq 1 \right\} } + \textrm{w}_t \left( \frac{U_{i-1,j,t} - U_{i,j,t}}{\Delta x} \right) \mathbbm{1}_{ \left\{ i-1 \geq 1 \right\} } \right. \label{eq: SPDE eq1} \\
&& - \tilde{\mu} U_{i,j,t} \left( \textrm{n}_t \mathbbm{1}_{ \left\{j = J \right\} } + \textrm{s}_t \mathbbm{1}_{ \left\{j = 1 \right\} } + \textrm{e}_t \mathbbm{1}_{ \left\{i = I \right\} } + \textrm{w}_t \mathbbm{1}_{ \left\{i = 1 \right\} } \right)  \Bigg) dt \\
&& + \tilde{\sigma} \Delta x dW_{i,j,t}  \nonumber \\
dQ_{i,j,t} &=& U_{i,j,t}dt \label{eq: SPDE eq2} \\
Y_{i,j,k} &=& Q_{i,j,t_k} +  \epsilon_{i,j,k}, \label{eq: SPDE eq3}
\end{eqnarray}

Now notice that for the easterly direction
\begin{eqnarray}
\lim_{\Delta x \to 0}\frac{U_{i+1,j,t} - U_{i,j,t}}{\Delta x} = \left( \frac{\partial U(x,t)}{ \partial x_1} \right)_{x_{i,j}}.
\end{eqnarray}
Similarly, we can compute this quantity for the other directions.

Next, consider the integral:
\begin{eqnarray}
\int_t^{t + \Delta t} \Delta x dW_{i,j,t} \sim \mathcal{N}(0,\Delta x^2 \Delta t) \qquad \forall \quad \Delta x, \Delta t \geq 0.
\end{eqnarray}
This is exactly equal to the definition of a Brownian motion in 2D space and time.

Hence it becomes evident that, when we look at the set of coupled stochastic differential equations given by (\ref{eq: SPDE model eq1}) away from the boundaries, this set can be interpreted as a finite difference discretization of the following stochastic partial differential equation (SPDE):
\begin{eqnarray}
d U(x,t) &=& \bar{v} \theta \nabla U(x,t) dt  + \sigma dW(x,t) \label{eq: SPDE interpretation},
\end{eqnarray}
where $W(x,t)$ is a Brownian motion in space and time, $\bar{v}$ is the cloud speed vector, and $\nabla$ is the partial derivative operator. Other SPDE models have originated following an analogous micro-to-macro approach (\cite{allen2008derivation}, \cite{hairer2009introduction}).

Some intuition about the SPDE (\ref{eq: SPDE interpretation}) can be gained by looking at its deterministic part, namely:
\begin{eqnarray}
d U(x,t) &=& \bar{v} \theta \nabla U(x,t) dt ,
\end{eqnarray}
which is a uni-direction wave equation, where the direction is determined by the cloud speed vector.

\section{Fitting Procedure} \label{sec: CTSM-R}

The estimation of parameters is carried out using the statistical software R (\cite{Rsoftware2014}) and in particular the package CTSM-R (Continuous Time Stochastic Modeling for R) (\cite{juhl2013ctsm}). The method is based on the Kalman filter for obtaining the likelihood. We provide a brief overview of the approach implemented and refer the reader to \cite{juhl2013ctsm} for more details on the approach.

The discretized SPDE (\ref{eq: SPDE eq1})-(\ref{eq: SPDE eq3}) is linear in the states $U_{i,j}$ and thus can be formulated as a linear SDE with linear observations
\begin{eqnarray}
   dU_{i,j,t} &=&  \mathbf A(\theta,t) \mathbf{U}_t dt + \Sigma(\theta) dW_{i,j,t} \label{eq: linear model eq1} \\ 
   Y_{k} &=& \mathbf C(\theta) \mathbf{U}_{t_k} +  \epsilon_{k} \label{eq: linear model eq2}, 
\end{eqnarray}
where $\mathbf A(\theta,t)$ is a time varying transition matrix, $\Sigma(\theta)$ is the diffusion matrix, $\mathbf C(\theta)$ determines how the states are observed and $e_k \sim \mathcal{N}(0,\sigma^2)$. The aim is to estimate the parameter vector $\theta$ in the model defined by the linear equations (\ref{eq: linear model eq1}) - (\ref{eq: linear model eq2}). The likelihood depends only on the one-step ahead prediction probability densities of the observations. A linear model driven by Gaussian diffusion results in a Gaussian process which is fully described by the mean and variance of the observations. We define them as
\begin{eqnarray}
\widehat{Y}_{k|k-1} &=& \mathbb{E}\left[Y_k | \mathcal{Y}_{k-1}, \theta \right]  \\
R_{k|k-1} &=& \mathbb{V} \left[Y_k | \mathcal{Y}_{k-1}, \theta \right],
\end{eqnarray}
where $\mathbb{E}\left[ \cdot \right] $ and $\mathbb{V} \left[ \cdot \right]$ denote the expectation and variance, respectively, and $\mathcal{Y}_{k-1} = \left\{ Y_0, \ldots, Y_{k-1} \right\}$. We can now define the innovation error as the difference between the observed and expected outcome:
\begin{eqnarray}
\epsilon_k = Y_k - \widehat{Y}_{k|k-1},
\end{eqnarray}
which will be used to compute the likelihood. We require in equations (\ref{eq: general model eq1}) - (\ref{eq: general model eq1}) that $g(U_t,t) = g(t)$  and $h(U_{t_k}, t_k, e_k)= h(U_{t_k}, t_k) +  e_k$, where $e_k \sim \mathcal{N}(0,\sigma^2)$. However these requirements can be alleviated to a large extent through transformations of the state equations (\cite{iversen2014probabilistic}, \cite{Moller2010Lamperti}) or transformations of the observations (\cite{box1964analysis}). For a system satisfying these conditions, the likelihood is given by
\begin{eqnarray}
L \left( \theta ; \mathcal{Y}_N \right) = \left( \prod_{k=1}^N{ \frac{ \exp \left( - \frac{1}{2} \epsilon_k^{\top} R_{k|k-1}^{-1} \epsilon_k \right) }{\sqrt{ \det\left( R_{k|k-1} \right)} \left( \sqrt{2 \pi} \right)^l } } \right) p(Y_0| \theta).
\end{eqnarray}
Here $l$ is the dimension of the sample space, thus the dimension of $Y_k$, $N$ is the number of observations, $(\cdot)^{\top}$ denotes the vector transpose and $p(Y_0| \theta)$ is the likelihood of seeing observation $Y_0$.

We are tracking changes in the observed power output, but by far the majority of the data is without cloud activity. To reduce the computational load, we extract $M = 12$ segments, each of 3 hours each with power measurements every 5 seconds to reduce the burden of the estimation process. The $M$ sets are from separate days spread out such that we have a sample day from each month of the year. Thus, the data sets can be assumed independent. The likelihood of $M$ independent sets of observations is
\begin{eqnarray}
L \left( \theta ; \mathbf{Y} \right) = \prod_{i=1}^M \left( \prod_{k=1}^{N_i}{ \frac{ \exp \left( - \frac{1}{2} {\epsilon^i_k}^{\top} {R^i_{k|k-1}}^{-1} {\epsilon^i_k} \right) }{\sqrt{ \det\left( R^i_{k|k-1} \right)} \left( \sqrt{2 \pi} \right)^l } } \right) p(Y^i_0| \theta) ,
\end{eqnarray}
where $\mathbf Y = \left[ \mathcal{Y}^1_{N_1}, \mathcal{Y}^2_{N_2}, \dots, \mathcal{Y}^M_{N_M} \right]$  is the combined set of observations and $N_i$ is the number of observations in each data set. We consider the logarithm of the likelihood function conditional on $\mathbf Y_0 = \left[ Y^1_0, Y^2_0, \dots, Y^M_0 \right]$, both for computational considerations and in order to deal with the fact that there are no observations prior to $Y_0$. This results in:
\begin{eqnarray}
 \log \left( L \left( \theta ; \mathbf Y_N \right|Y_0) \right) &=& - \frac{1}{2} \sum_{i=1}^M \sum_{k=1}^{N_i}{ \left( \log (\det (R^i_{k|k-1}))\! +\! {\epsilon^i_k}^{\top} {R^i_{k|k-1}}^{-1} \epsilon^i_k  \right)} \nonumber \\
&& - \log (2 \pi) \frac{l \sum_{i=1}^M N_i}{2}. \label{eq:log-likelihood} 
\end{eqnarray}

The parameter vector $\theta$ enters the log-likelihood function~(\ref{eq:log-likelihood}) through $\epsilon^i_k$ and $R^i_{k|k-1}$. An estimate of the parameters in the model can now be obtained by maximizing~(\ref{eq:log-likelihood}), i.e.,
\begin{equation}
\hat{\theta}=\textrm{arg} \max_{\theta \in \Theta} \left( \log \left( L \left( \theta ; \mathbf{Y}_N \right| \mathbf Y_0) \right) \right), \label{eq:optim}
\end{equation}
where $\Theta$ is the feasible parameter space. A thorough introduction to parameter estimation and filtering is found in \cite{jazwinski2007stochastic}. We note that the likelihood function is optimized for the one-step-ahead residuals. To estimate SDE models using a multi-horizon approach, we refer the interested reader to \cite{Moller2013Probabilistic}.

\subsection{Speeding up the estimation process}

We use a large amount of data during the estimation procedure which naturally slows down the computation. Since the likelihood of the 12 independent sets of observations is simply the product of the likelihood of each of the sets, we can evaluate these likelihoods in parallel.
Furthermore, the optimization problem (\ref{eq:optim}) is solved using a quasi Newton algorithm, where the gradient of the log likelihood (\ref{eq:log-likelihood}) is determined by a finite difference scheme. This means evaluating (\ref{eq:log-likelihood}) at several independent points in the parameter space $\Theta$. Thus, the computation of the gradient can also be parallelized.

A server with 2x12 cores AMD Opteron 6168 CPUs was used for estimation and prediction. We use OpenMP and nested parallelism to maximize the use of the server. The log likelihood (\ref{eq:log-likelihood}) is always computed in parallel using 12 threads. When computing the gradient, two evaluations of the log likelihood were allowed simultaneously, and in so doing all the available 24 cores were used. We achieved a total speedup of 15.6x.

\section{Nowcasting at Copper Mountain First Solar} \label{sec: Nowcasting at Copper Mountain}

In this section we look at the specific problem of nowcasting the power output of the Copper Mountain Solar facility. We fit the model on a training data set consisting of data from 12 days, one from each month of the year, where we select 3 hours around noon. We select only days where there are actually observed clouds as we propose to model the cloud dynamics. Similarly, we select a test data-set not overlapping with the training data set. We first run the estimation procedure described in Section \ref{sec: CTSM-R} on the model described in Section \ref{sec: Solar Forecasting Model}. As this is computationally intensive, we estimate the parameters on a cutout of 5 by 7 inverters to allow for a timely estimation procedure. The parameters found here are then used to define the full model spanning 5 by 14 inverters. The parameter estimates obtained are shown in Table \ref{tab:ParameterEstimates}.

\begin{table}[ht]
\small
\begin{center}
    \begin{tabular}{cccc}
    \toprule
    $\hat{\theta}$ & $\hat{\mu}$   & $\hat{\sigma}$    &  $\hat{\sigma}_{\epsilon}$\\ 
        \midrule
    0.0631 &  0.703  &  0.00865  &  $10^{-10}$ \\
   \bottomrule
\end{tabular}
\end{center}
 \vspace{-10pt}
  \caption{\emph{Parameter estimates for the proposed model }}
\label{tab:ParameterEstimates}
\end{table}

\begin{figure}[!ht]
\begin{center}
  \includegraphics[width=0.95\textwidth]{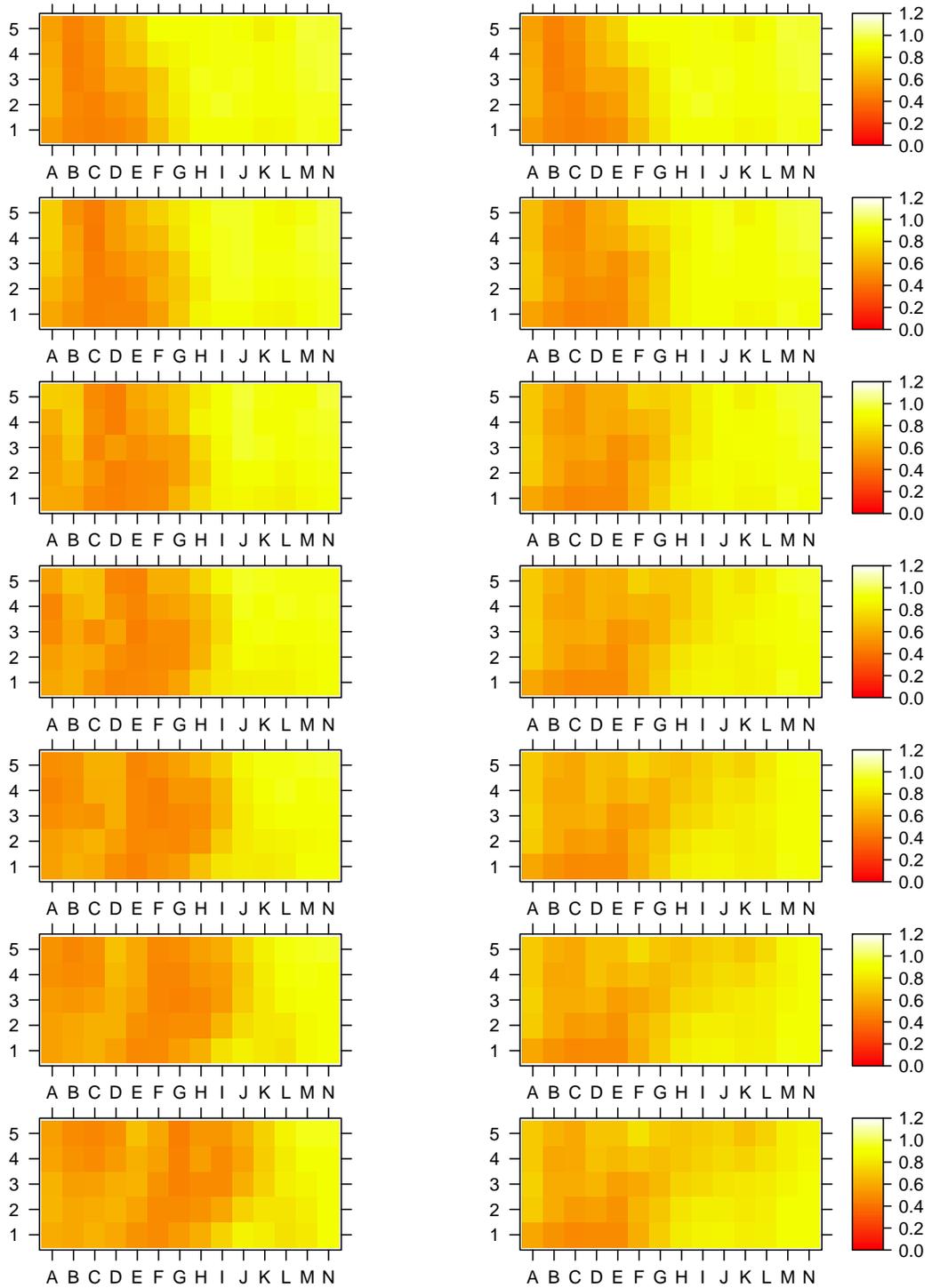}\\
    \vspace{-10pt}
  \caption{\emph{The observed power generation (left) and the predicted power generation (right) from 0 to 60 seconds in 10 second increments. The x-axis corresponds to the East-West axis and the y-axis corresponds to the North-South axis.}}\label{fig:SpatialExample}
\end{center}
\vspace{-10pt}
\end{figure}

Figure \ref{fig:SpatialExample} displays the actual power output of the inverters along with the predicted power for different horizons. It can be observed that we manage to track cloud movement through the system as regions with lower power output. However, for longer horizons the predicted power may vary to a large degree from the observed power. This can be explained by several facts: first, that the clouds that caused the drop in power were not observed by the upwind inverters at the time when the forecast was issued. Second, there might be some smaller errors in the estimation of the cloud speed vector. These errors compound to produce larger errors for larger lead times. Third, looking at the observations in the left panels, it seems that the actual structure of the cloud actually changes over time. This may, in part, be due to the spatial resolution of the observations and, in part, due to real changes of the cloud structure.

\begin{figure}[!ht]
\begin{center}
  \includegraphics[width=0.9\textwidth]{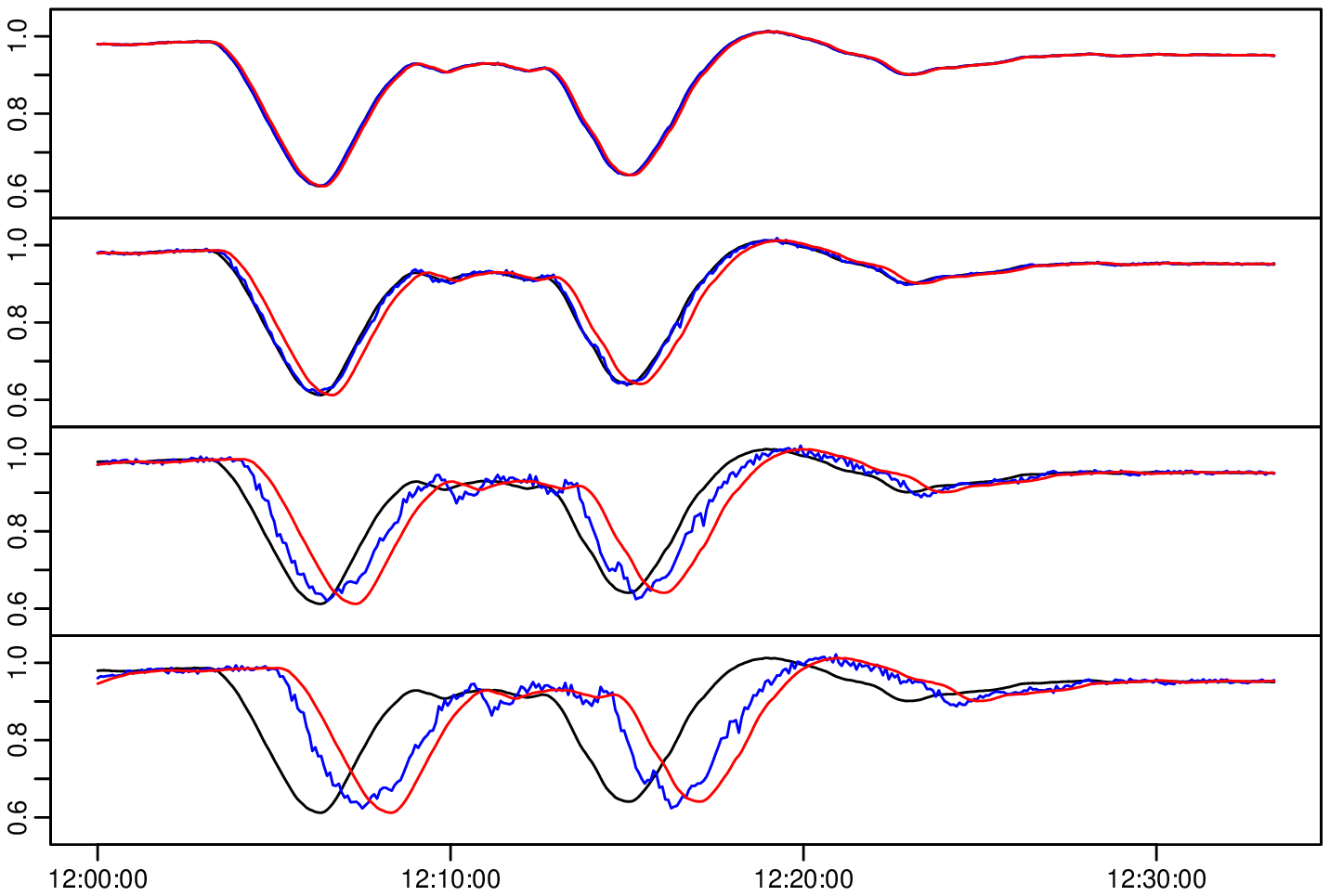}\\
    \vspace{-10pt}
  \caption{\emph{The observed normalized total power in black, along with the (from top to bottom) 5, 20, 60 and 120 sec ahead forecasts from the model in blue and the persistence forecast in red. The x-axis is time in hours, minutes and seconds, the y-axis is normalized power. }}\label{fig:TotalPowerExample}
\end{center}
\vspace{-10pt}
\end{figure}

In Figure \ref{fig:TotalPowerExample} the normalized total power output of the solar field is shown together with predictions issued for different lead times. We see that for 5- and 20-second horizons the model seems to be successful at predicting the output power. For longer horizons we can observe a ``lagging'' behavior. This is caused by the fact that the clouds causing the power drop (or increase) had not yet begun to enter into the system (leave the system) and therefore, to be detected. As a result, their future effects are not anticipated by the forecasts. Furthermore, the predictions become less smooth as we predict for longer lead times. This is caused by the predictions for total power being based on fewer actual observations, since the influence of many inverters is propagated out of the system. There is also an error propagation, where small errors accumulate over time.

Second, we compare the model proposed in this paper with state-of-the-art models for spatio-temporal solar power forecasting. We compare the different models on several horizons to better understand the specific characteristics of each . The benchmarks include a cloud speed persistence model (as defined in \cite{lipperheide2015embedded}), which propagates the power production along the cloud speed vector. Another benchmark is the ramp speed persistence model (also in \cite{lipperheide2015embedded}), where the change in power is assumed to stay constant for the near-term future. A third benchmark is an auto-regressive model defined as:
\begin{eqnarray}
Y_{k} = \psi_0 +  \sum_{i=1}^{p}{ \psi_i Y_{k-i}} + \epsilon_{k}, \qquad \textrm{where } \epsilon_{k} \sim \mathcal{N}(0,\sigma^2),
\end{eqnarray}
$\psi_i$ are the auto-regressive parameters of the model and $p$ defines the number of lags included.

In Table \ref{tab:Scores} the proposed model is compared with the benchmarks in terms of skill scores against the persistence benchmark for the root mean squared error (RMSE) and the mean absolute error (MAE) of total power production. The persistence is the lagged value of the observations, here the lag is given as the forecast horizon. This skill score is computed as: 
\begin{eqnarray}
\textrm{SS} = 1 - \frac{S_{\textrm{forecast}}}{S_{\textrm{ref}}},
\end{eqnarray}
where $\textrm{SS}$ is the skill score for the forecast score, $S_{\textrm{forecast}}$, against the reference score, $S_{\textrm{ref}}$ , obtained from the reference model (In this application the reference model is the persistence).

For benchmarks that produce predictive densities, we compare in terms of continuous ranked probability score (CRPS) (as defined in \cite{gneiting2007strictly}) to evaluate the probabilistic properties of the predictions.

\begin{table}[ht]
\small
\begin{center}
    \begin{tabular}{l|rrrrrrr}
    \toprule
               & Cloud Speed & Ramp Speed & Auto- & Model  \\ 
 Score       & Persistence & Persistence & Regressive & (\ref{eq: SPDE model eq1})-(\ref{eq: SPDE model eq3}) \\    
   \midrule
  $\textrm{RMSE}_5$ 		& 0.334  & 0.612  & 0.464 & \textbf{0.636} \\
  $\textrm{RMSE}_{20}$      & 0.289  & 0.284  & 0.319 & \textbf{0.523} \\
  $\textrm{RMSE}_{60}$      & 0.168  & -0.203 & 0.113 & \textbf{0.254} \\
  $\textrm{RMSE}_{120}$     & 0.062  & -0.434 & 0.039 & \textbf{0.097} \\
     \midrule
  $\textrm{MAE}_5$ 		    & 0.258  & 0.597  & 0.431  & \textbf{0.612} \\
  $\textrm{MAE}_{20}$       & 0.213  & 0.301  & 0.280  & \textbf{0.497} \\
  $\textrm{MAE}_{60}$       & 0.136  & -0.145 & 0.045  & \textbf{0.246} \\
  $\textrm{MAE}_{120}$      & 0.048  & -0.396 & -0.064 & \textbf{0.096} \\
     \midrule
  $\textrm{CRPS}_5$ 	     & $-$ & $-$  & 0.00262  & \textbf{0.00131} \\
  $\textrm{CRPS}_{20}$       & $-$ & $-$  & 0.00982  & \textbf{0.00666} \\
  $\textrm{CRPS}_{60}$       & $-$ & $-$  & 0.02886  & \textbf{0.02455} \\
  $\textrm{CRPS}_{120}$      & $-$ & $-$  & 0.04883  & \textbf{0.04675} \\
   \bottomrule 
\end{tabular}
\end{center}
 \vspace{-10pt}
  \caption{\emph{The MAE skill score, RMSE skill score and CRPS for benchmarks as well as for the proposed model for horizons of 5, 20, 60 and 120 seconds.}}
\label{tab:Scores}
\end{table}

In Table \ref{tab:Scores} we see that the coupled SDE model (\ref{eq: SPDE model eq1})--(\ref{eq: SPDE model eq3}) outperforms all benchmarks on all horizons in terms of all the proposed scores.

\begin{figure}[!ht]
\begin{center}
  \vspace{-50pt}
  \includegraphics[width=0.90\textwidth]{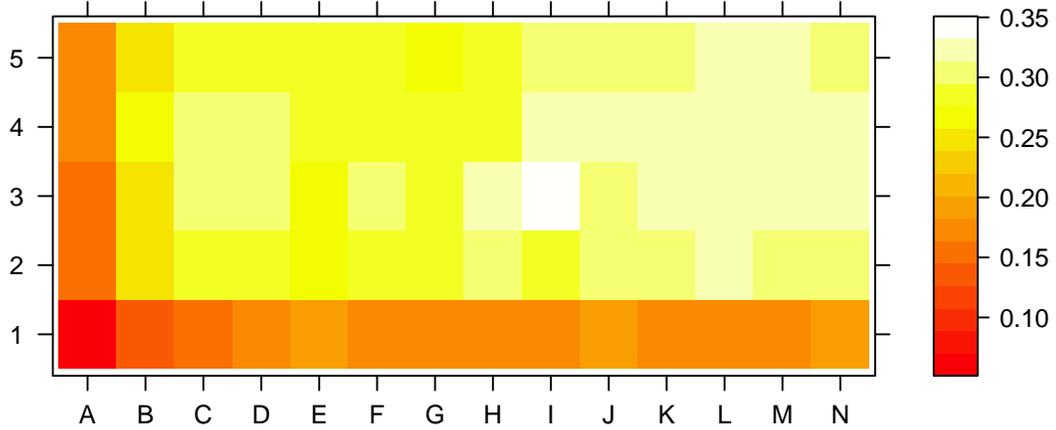}\\
    \vspace{-50pt}
  \caption{\emph{The RMSE skill score for 20 seconds ahead forecasts computed for individual inverters.}}\label{fig:SpatialRMSE}
\end{center}
\vspace{-10pt}
\end{figure}

The scores in Table \ref{tab:Scores} are computed on the basis of total output power. However, as the proposed model also captures the dynamics of each individual inverter,  we might well evaluate the predictive performance of the individual inverters. This is done with respect to the RMSE skill score for 20 seconds ahead in Figure \ref{fig:SpatialRMSE}. As this is a skill-score, higher score values are better. As seen in Figure \ref{fig:SpatialRMSE} the inverters that perform the poorest are located on the southern and western limits of the solar plant. Investigating this phenomenon we find that prevailing winds are south-westerly. We would expect the up-wind inverters to perform worse compared to down wind inverters due to the influx of clouds. Thus the findings from our model are in accordance with what would be expected.

\begin{figure}[!ht]
\begin{center}
  \vspace{-50pt}
  \includegraphics[width=0.90\textwidth]{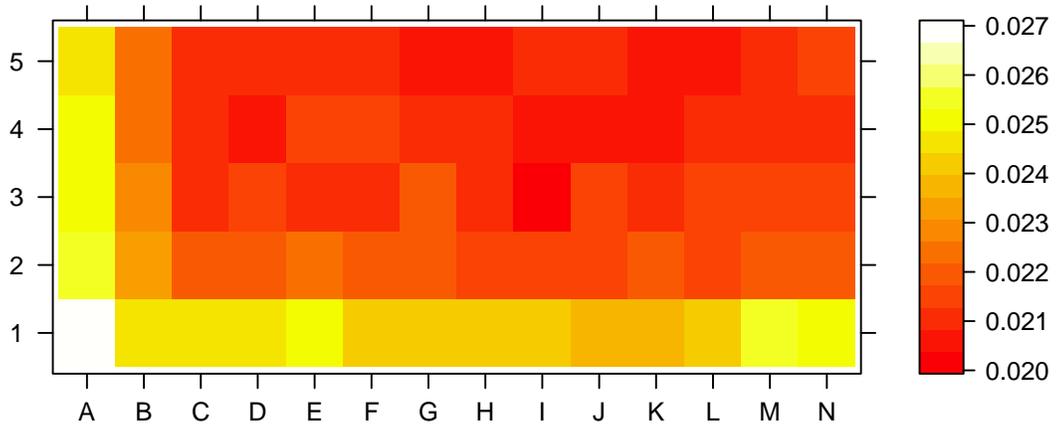}\\
    \vspace{-50pt}
  \caption{\emph{The CRPS for 20 seconds ahead forecasts computed for individual inverters. The x-axis corresponds to the East-West axis and the y-axis corresponds to the North-South axis.}}\label{fig:SpatialCRPS}
\end{center}
\vspace{-10pt}
\end{figure}

Figure \ref{fig:SpatialCRPS} is analogous to Figure \ref{fig:SpatialRMSE} but in terms of the CRPS. As opposed to Figure \ref{fig:SpatialRMSE}, lower CRPS values are better. Again we see the better performance in the interior of the solar power plant.

The proposed spatio-temporal model outperforms persistence as well as all the proposed state-of-the-art benchmarks. A spatial understanding of the dynamics not only allows for spatio-temporal predictions, but also allows us to better predict the aggregate power production. Furthermore, we note that the performance of the benchmarks that are used here is similar to the performance of the benchmarks found in \cite{lipperheide2015embedded}. It is crucial to stress the importance of the inputs into the model being the cloud speed vector obtained through the approach in \cite{bosch2013cloud}. This cloud speed vector has crucial importance for the correct propagation of the irradiance field. As mentioned in \cite{bosch2013cloud}, estimating the correct cloud speed vector is not a simple task. This is, in part, due to the granularity of the spatial observations and, in part, due to the deformation of the clouds. Even small errors in the cloud speed vector lead to serious errors when the forecast horizon is increased. In \cite{bosch2013cloud} it is also clearly stated that there is a large degree of uncertainty related to this clouds speed vector.

\section{Concluding Remarks} \label{sec: Conclusion}

Spatio-temporal problems arise in many fields ranging from physics to ecology. The processes that drive these systems can be quite complicated. However, there are often scientific theories that explain the local behavior of the system, e.g., mass and energy balances or the migration of animals. Although, generally, traditional statistical methods are not well-suited to model such processes, the modeling framework that we propose in this paper, based on coupled stochastic differential equations, proves to perform satisfactorily. Coupled stochastic differential equations have the capacity to capture dynamics where the structure of the governing stochastic partial differential equation is not well known and they can be used to identify a possible candidate, as we showed here. Furthermore, this approach bridges the gap between spatio-temporal models that are driven by physics and those which are driven by data. A key feature is that the model framework reduces the parameter dimension of the spatio-temporal problem and thereby facilitates parameter estimation and efficient computation. In this paper the methodology for building spatio-temporal models is applied to forecasting solar power generation at a solar power facility.

The spatio-temporal forecast model proposed in this paper outperforms state-of-the-art benchmarks on all horizons while also being able to provide scenarios, covariances and predictive densities. Understanding the spatial dynamics not only allows for spatio-temporal predictions but also allows us to produce higher quality predictions for aggregate power production. The model generates predictions swiftly and as such, could be run online. Thus, we produce a methodology for predicting large ramp events 30-60 seconds ahead in time (depending on cloud speed and direction) providing forecast users with an early warning to scramble alternative power generation.

The application to solar power forecasting that we introduce in this paper assumes the same local dynamics for every location. This is not, however, a requirement. Distinct inputs could be present for each specific location, distinguishing the dynamics at different grid points. Also, if this framework were to be applied to forecast power generation from distributed solar power in large urban areas, the grid would change and there may be distinct features in space to consider. A clear conclusion from the results shown in Figure \ref{fig:SpatialCRPS} is that the model could be extended to have irradiance sensors away from the solar power facility in order to increase forecast performance and to extend the forecast horizon.

The spatio-temporal model considered in this paper is of a particularly simple structure, with regular grid spacing in all spatial dimensions. Nonetheless, this is not a prerequisite for applying a similar model to a non-regular grid and as such, more research efforts can be dedicated at adapting this approach to irregular grids. Future work could also include using the sparse structure of the coupled stochastic differential equations to further reduce computational time. This sparsity is caused by the very nature of the approach, where only locations that are adjacent to each other interact.

\section*{Acknowledgments}
DSF (Det Strategiske Forskningsr\aa d) is to be acknowledged for partly
funding the work of Emil B. Iversen, Juan M. Morales and Henrik Madsen through the Ensymora project (no. 10-093904/DSF). Furthermore, Juan M. Morales and Henrik Madsen are partly funded by the Research Centre CITIES (no. 1035-00027B), which is also supported by DSF (Det Strategiske Forskningsr\aa d). The work was completed as part of a research stay at the University of California, San Diego, made by Emil B. Iversen.


\section*{References}
 \bibliographystyle{apalike}
\bibliography{SolarPlantJASA}

\begin{thebibliography}{}

\bibitem[Adomian, 1988]{adomian1988nonlinear}
Adomian, G. (1988).
\newblock {\em Nonlinear Stochastic Systems Theory and Application to Physics},
  volume~46.
\newblock Springer.

\bibitem[Allen, 2008]{allen2008derivation}
Allen, E.~J. (2008).
\newblock Derivation of stochastic partial differential equations.
\newblock {\em Stochastic Analysis and Applications}, 26(2):357--378.

\bibitem[{\AA}str{\"o}m and Bell, 2000]{aastrom2000drum}
{\AA}str{\"o}m, K.~J. and Bell, R.~D. (2000).
\newblock Drum-boiler dynamics.
\newblock {\em Automatica}, 36(3):363--378.

\bibitem[Bj{\"o}rk, 2009]{bjork2009arbitrage}
Bj{\"o}rk, T. (2009).
\newblock {\em Arbitrage Theory in Continuous Time}.
\newblock Oxford Finance Series. OUP Oxford.

\bibitem[Bosch and Kleissl, 2013]{bosch2013cloud}
Bosch, J. and Kleissl, J. (2013).
\newblock Cloud motion vectors from a network of ground sensors in a solar
  power plant.
\newblock {\em Solar Energy}, 95:13--20.

\bibitem[Bowman, 2007]{bowman2007spatiotemporal}
Bowman, F.~D. (2007).
\newblock Spatiotemporal models for region of interest analyses of functional
  neuroimaging data.
\newblock {\em Journal of the American Statistical Association},
  102(478):442--453.

\bibitem[Box and Cox, 1964]{box1964analysis}
Box, G.~E. and Cox, D.~R. (1964).
\newblock An analysis of transformations.
\newblock {\em Journal of the Royal Statistical Society. Series B
  (Methodological)}, pages 211--252.

\bibitem[Cressie and Wikle, 2011]{cressie2011statistics}
Cressie, N. and Wikle, C.~K. (2011).
\newblock {\em Statistics for spatio-temporal data}.
\newblock John Wiley \& Sons.

\bibitem[Gneiting and Raftery, 2007]{gneiting2007strictly}
Gneiting, T. and Raftery, A. (2007).
\newblock Strictly proper scoring rules, prediction, and estimation.
\newblock {\em Journal of the American Statistical Association},
  102(477):359--378.

\bibitem[Hairer, 2009]{hairer2009introduction}
Hairer, M. (2009).
\newblock An introduction to stochastic {PDEs}.
\newblock {\em arXiv preprint arXiv:0907.4178}.

\bibitem[Huang and Garcia, 1998]{huang1998herschel}
Huang, X. and Garcia, M.~H. (1998).
\newblock A {H}erschel--{B}ulkley model for mud flow down a slope.
\newblock {\em Journal of fluid mechanics}, 374:305--333.

\bibitem[{Iversen} et~al., 2014]{iversen2014probabilistic}
{Iversen}, E.~B., {Morales}, J.~M., {M{\o}ller}, J.~K., and {Madsen}, H.
  (2014).
\newblock {Probabilistic Forecasts of Solar Irradiance by Stochastic
  Differential Equations}.
\newblock {\em Environmetrics}, 25(3):152--164.

\bibitem[Jazwinski, 2007]{jazwinski2007stochastic}
Jazwinski, A.~H. (2007).
\newblock {\em Stochastic Processes and Filtering Theory}.
\newblock Courier Dover Publications.

\bibitem[Johnson, 2012]{johnson2012numerical}
Johnson, C. (2012).
\newblock {\em Numerical solution of partial differential equations by the
  finite element method}.
\newblock Courier Dover Publications.

\bibitem[Juhl et~al., 2013]{juhl2013ctsm}
Juhl, R., Kristensen, N.~R., Bacher, P., Kloppenborg, J., and Madsen, H.
  (2013).
\newblock {CTSM-R} user guide.
\newblock {\em Technical University of Denmark. Available at
  \url{http://ctsm.info/pdfs/userguide.pdf}}.

\bibitem[Lazer et~al., 2009]{lazer2009life}
Lazer, D., Pentland, A.~S., Adamic, L., Aral, S., Barabasi, A.~L., Brewer, D.,
  Christakis, N., Contractor, N., Fowler, J., Gutmann, M., et~al. (2009).
\newblock Life in the network: {T}he coming age of computational social
  science.
\newblock {\em Science}, 323(5915):721.

\bibitem[Lemos and Sans{\'o}, 2009]{lemos2009spatio}
Lemos, R.~T. and Sans{\'o}, B. (2009).
\newblock A spatio-temporal model for mean, anomaly, and trend fields of
  {N}orth {A}tlantic sea surface temperature.
\newblock {\em Journal of the American Statistical Association},
  104(485):5--18.

\bibitem[Lipperheide et~al., 2015]{lipperheide2015embedded}
Lipperheide, M., Bosch, J., and Kleissl, J. (2015).
\newblock Embedded nowcasting method using cloud speed persistence for a
  photovoltaic power plant.
\newblock {\em Solar Energy}, 112:232--238.

\bibitem[Mikosch, 1998]{mikosch1998elementary}
Mikosch, T. (1998).
\newblock {\em Elementary Stochastic Calculus: With Finance in View}, volume~6.
\newblock World Scientific.

\bibitem[M{\"o}ller et~al., 2013]{moller2013multivariate}
M{\"o}ller, A., Lenkoski, A., and Thorarinsdottir, T.~L. (2013).
\newblock Multivariate probabilistic forecasting using ensemble {B}ayesian
  model averaging and copulas.
\newblock {\em Quarterly Journal of the Royal Meteorological Society},
  139(673):982--991.

\bibitem[M{\o}ller and Madsen, 2010]{Moller2010Lamperti}
M{\o}ller, J. and Madsen, H. (2010).
\newblock From state dependent diffusion to constant diffusion in stochastic
  differential equations by the {L}amperti transform.
\newblock Technical report, Technical University of Denmark.

\bibitem[M{\o}ller et~al., 2013]{Moller2013Probabilistic}
M{\o}ller, J., Pinson, P., and Madsen, H. (2013).
\newblock Probabilistic forecasts of wind power generation by stochastic
  differential equation models.
\newblock Technical report, Technical University of Denmark.

\bibitem[Morales et~al., 2014]{morales2014decision}
Morales, J.~M., Conejo, A.~J., Madsen, H., Pinson, P., and Zugno, M. (2014).
\newblock {\em Integrating Renewables in Electricity Markets -- Operational
  Problems}, volume 205 of {\em International Series in Operations Research \&
  Management Science}.
\newblock Springer, New York.

\bibitem[{\O}ksendal, 2010]{oksendal2010stochastic}
{\O}ksendal, B. (2010).
\newblock {\em Stochastic Differential Equations: An Introduction with
  Applications}.
\newblock Universitext (1979). Springer.

\bibitem[Pinson et~al., 2013]{pinson2013wind}
Pinson, P. et~al. (2013).
\newblock Wind energy: Forecasting challenges for its operational management.
\newblock {\em Statistical Science}, 28(4):564--585.

\bibitem[{R Core Team}, 2014]{Rsoftware2014}
{R Core Team} (2014).
\newblock {\em R: A Language and Environment for Statistical Computing}.
\newblock R Foundation for Statistical Computing, Vienna, Austria.

\bibitem[Richardson et~al., 2010]{richardson2010bayesian}
Richardson, S., Bottolo, L., and Rosenthal, J.~S. (2010).
\newblock Bayesian models for sparse regression analysis of high dimensional
  data.
\newblock {\em Bayesian Statistics}, 9:539--569.

\bibitem[Robert and Casella, 2004]{robert2004monte}
Robert, C.~P. and Casella, G. (2004).
\newblock {\em Monte Carlo statistical methods}, volume 319.
\newblock Citeseer.

\bibitem[Robertson et~al., 2010]{robertson2010review}
Robertson, C., Nelson, T.~A., MacNab, Y.~C., and Lawson, A.~B. (2010).
\newblock Review of methods for space--time disease surveillance.
\newblock {\em Spatial and Spatio-temporal Epidemiology}, 1(2):105--116.

\bibitem[Sigrist et~al., 2014]{sigrist2014stochastic}
Sigrist, F., K{\"u}nsch, H.~R., and Stahel, W.~A. (2014).
\newblock Stochastic partial differential equation based modelling of large
  space--time data sets.
\newblock {\em Journal of the Royal Statistical Society: Series B (Statistical
  Methodology)}.

\bibitem[Smith, 1965]{smith1965numerical}
Smith, G.~D. (1965).
\newblock Numerical solution of partial differential equations.

\bibitem[Tastu et~al., 2013]{tastu2013space}
Tastu, J., Pinson, P., and Madsen, H. (2013).
\newblock Space-time scenarios of wind power generation produced using a
  {G}aussian copula with parametrized precision matrix.
\newblock Technical report, Technical University of Denmark.

\bibitem[Uppala et~al., 2005]{uppala2005era}
Uppala, S.~M., K{\aa}llberg, P., Simmons, A., Andrae, U., Bechtold, V.,
  Fiorino, M., Gibson, J., Haseler, J., Hernandez, A., Kelly, G., et~al.
  (2005).
\newblock The {ERA-40} re-analysis.
\newblock {\em Quarterly Journal of the Royal Meteorological Society},
  131(612):2961--3012.

\bibitem[Van~Kampen, 1992]{van1992stochastic}
Van~Kampen, N.~G. (1992).
\newblock {\em Stochastic Processes in Physics and Chemistry}, volume~1.
\newblock Access Online via Elsevier.

\bibitem[Van~Maanen and Xu, 2003]{van2003modelling}
Van~Maanen, A. and Xu, X.-M. (2003).
\newblock Modelling plant disease epidemics.
\newblock {\em European Journal of Plant Pathology}, 109(7):669--682.

\bibitem[Versini, 2012]{versini2012use}
Versini, P.-A. (2012).
\newblock Use of radar rainfall estimates and forecasts to prevent flash flood
  in real time by using a road inundation warning system.
\newblock {\em Journal of Hydrology}, 416:157--170.

\bibitem[Waller et~al., 1997]{waller1997hierarchical}
Waller, L.~A., Carlin, B.~P., Xia, H., and Gelfand, A.~E. (1997).
\newblock Hierarchical spatio-temporal mapping of disease rates.
\newblock {\em Journal of the American Statistical association},
  92(438):607--617.

\bibitem[Xu et~al., 2005]{xu2005kernel}
Xu, K., Wikle, C.~K., and Fox, N.~I. (2005).
\newblock A kernel-based spatio-temporal dynamical model for nowcasting weather
  radar reflectivities.
\newblock {\em Journal of the American Statistical Association},
  100(472):1133--1144.

\bibitem[Yang et~al., 2015]{yang2014multitime}
Yang, C., Thatte, A., and Xie, L. (2015).
\newblock Multitime-scale data-driven spatio-temporal forecast of photovoltaic
  generation.
\newblock {\em Sustainable Energy, IEEE Transactions on}, 6(1):104--112.

\bibitem[Yang et~al., 2014]{yang2014solar}
Yang, D., Dong, Z., Reindl, T., Jirutitijaroen, P., and Walsh, W.~M. (2014).
\newblock Solar irradiance forecasting using spatio-temporal empirical kriging
  and vector autoregressive models with parameter shrinkage.
\newblock {\em Solar Energy}, 103:550--562.

\end{thebibliography}

\end{document}